\documentclass[12pt]{iopart}
\usepackage{graphicx}
\usepackage{dcolumn}
\usepackage{bm}
\usepackage[utf8]{inputenc}
\usepackage{graphicx}
\usepackage{color}
\usepackage{amssymb}
\usepackage{subfigure}
\usepackage{epsfig}
\usepackage{float}
\usepackage{lipsum}
\usepackage{cases}
\usepackage{appendix}
\usepackage{cancel}
\usepackage{soul}
\usepackage{cite}

\newcommand{\jj}{\boldsymbol{j}}

\newcommand{\D}{\boldsymbol{\mathrm{D}}}

\newcommand{\Gama}{\boldsymbol{\Gamma}}

\newcommand{\M}{\boldsymbol{\mathrm{M}}}

\newcommand{\B}{\boldsymbol{B}}

\newcommand{\T}{\boldsymbol{\mathrm{T}}}
\newcommand{\A}{\boldsymbol{\mathrm{A}}}
\newcommand{\1}{\boldsymbol{1}}
\newcommand{\LL}{\boldsymbol{\mathrm{L}}}
\newcommand{\Lambdaa}{\boldsymbol{\Lambda}}

\newcommand{\CC}{\boldsymbol{\mathcal{C}}}

\newcommand{\kk}{\boldsymbol{k}}

\newcommand{\nn}{\boldsymbol{n}}
\newcommand{\J}{\boldsymbol{J}}

\newcommand{\dif}{\mathop{}\!\mathrm{d}}
\newcommand{\Det}{\mathop{}\!\mathrm{Det}}

\newcommand{\diag}{\mathop{}\!\boldsymbol{\mathrm{diag}}}
\newcommand{\uu}{\boldsymbol{u}}
\newcommand{\phii}{\boldsymbol{\Phi}}

\newcommand{\vv}{\boldsymbol{v}} 
\newcommand{\ww}{\boldsymbol{w}}
\newcommand{\rr}{\boldsymbol{r}}

\newcommand{\0}{\boldsymbol{0}}

\newcommand{\xxi}{\boldsymbol{\xi}}

\newcommand{\eeta}{\boldsymbol{\eta}}

\newcommand{\R}{\boldsymbol{\mathrm{R}}}

\newcommand{\cmnt}[1]{{\color{black}{ #1}}} 

\newcommand{\eqref}[1]{({\ref{#1}})}

\begin{document}

\title[]{Correlations in multithermostat Brownian systems with Lorentz force}

\author{Iman Abdoli$^1$, Erik Kalz$^1$, Hidde D Vuijk$^1$, Ren\'e Wittmann$^2$, Jens-Uwe Sommer$^{1,3}$, Joseph M Brader$^{4}$ and Abhinav Sharma$^{1,3}$}

\address{$^1$ Leibniz-Institut  f\"ur Polymerforschung Dresden, Institut Theorie der Polymere, 01069 Dresden, Germany}
\address{$^2$ Institut f\"ur Theoretische Physik II, Weiche Materie, Heinrich-Heine-Universit\"at D\"usseldorf, 40225 D\"usseldorf, Germany}
\address{$^3$ Technische Universit\"at Dresden, Institut f\"ur Theoretische Physik, 01069 Dresden, Germany}
\address{$^4$ Department de Physique, Universit\'e de Fribourg, CH-1700 Fribourg, Switzerland}
\ead{sharma@ipfdd.de}
\vspace{10pt}

\begin{abstract}
We study the motion of a Brownian particle subjected to Lorentz force due to an external magnetic field. Each spatial degree of freedom of the particle is coupled to a different thermostat. We show that the magnetic field results in correlation between different velocity components in the stationary state. Integrating the velocity autocorrelation matrix, we obtain the diffusion matri that enters the Fokker-Planck equation for the probability density. The eigenvectors of the diffusion matrix do not align with the temperature axes. As a consequence the Brownian particle performs spatially correlated diffusion. We further show that in the presence of an isotropic confining potential, an unusual, flux-free steady state emerges which is characterized by a non-Boltzmann density distribution, which can be rotated by reversing the magnetic field. The nontrivial steady state properties of our system result from the Lorentz force induced coupling of the spatial degrees of freedom which cease to exist in equilibrium corresponding to a single-temperature system. 
\end{abstract}

%
\vspace{2pc}

%
%
%
%

\section{Introduction}

The trajectory of a charged particle is curved by the Lorentz force due to an external magnetic field.
Since the field does not perform work on the particle, the equilibrium properties of the system are unaffected by the applied magnetic field. However, the Lorentz force affects the dynamics of the system. For instance, it is well known that in diffusive systems Lorentz force reduces the diffusion coefficient of the particle in the plane perpendicular to the magnetic field, whereas the diffusion along the field is unaffected~\cite{balakrishnan2008elements,vuijk2019anomalous,vuijk2019effect}. Even in overdamped systems, the hallmark signature of Lorentz force -- deflection of trajectories in the direction perpendicular to the velocity -- is manifested. This deflection gives rise to additional Lorentz fluxes perpendicular to the typical diffusive fluxes~\cite{chun2018emergence, abdoli2020nondiffusive, abdoli2020stationary}. 

The Lorentz fluxes, akin to the diffusive Hall effect, generate dynamics which are fundamentally different from a purely diffusive system. We have recently shown that by driving the system into a nonequilibrium stationary state one preserves the unusual features of the dynamics under Lorentz force. 
Considering an internally driven system of active Brownian particles subjected to a spatially inhomogeneous Lorentz force, we showed that the resulting nonequilibrium steady state is characterized by
an inhomogeneous density distribution and macroscopic fluxes ~\cite{vuijk2019lorenz}. In another study, we used stochastic resetting~\cite{evans2011diffusion, evans2011optimal, evans2014diffusion} to drive a (passive) Brownian system into a nonequilibrium steady state with a non-Gaussian probability distribution and Lorentz fluxes~\cite{abdoli2020stationary}. 

The unusual properties of the stationary regime have their origin in the fact that the Lorentz force mixes different velocity components of the particle which results in coupling of the spatial degrees of freedom~\cite{voropajeva2008correlation}. If one now considers that each spatial degree of freedom is coupled to a different thermostat, an interesting steady state may be envisaged in the presence of Lorentz force. Here, we follow this approach to drive the system into a nonequilibrium steady state. The notion of physical systems characterized by two different temperatures was originally employed in neural networks and spin glasses with partially annealed disorder~\cite{penney1993coupled, dotsenko1994partial, feldman1994partially}. \cmnt{Multiple thermostats are also used in the models for mixtures of active and passive particles, in which  the active species is either coupled to a higher temperature than the passive one~\cite{grosberg2015nonequilibrium, weber2016binary, tanaka2017hot, ilker2020phase, chari2019scalar} or specified by a colored noise~\cite{chaki2018entropy, chaki2019effects,chaki2020escape,PhysRevE.102.012609,wittmann2017effective1, wittmann2017effective2}. }


Recently, the notion of multiple thermostats was applied on a single-particle level: each spatial degree of freedom of a Brownian particle was coupled to a different temperature~\cite{dotsenko2013two, murashita2016overdamped,holubec2017thermal,
nascimento2020memory, nascimento2020stationary}. In Ref.~\cite{dotsenko2013two} it was shown that in the presence of an anisotropic potential, the two-dimensional system settled into a nonequilibrium stationary state, characterized by the presence of space-dependent particle currents and a non-Boltzmann density distribution. The emergence of these particle currents is due to two broken symmetries: a different temperature in each spatial degree of freedom and a mismatch between the temperature axes, and the potential principal axes \cite{mancois2018two}. Although the interest in such systems remains primarily theoretical, a possible experimental realisation, based on cold atoms, has been suggested in Ref.~\cite{mancois2018two}, in which, by detuning laser intensity along the two axes, one obtains two different temperatures in the optical trap. \cmnt{Including an external magnetic field in such a system is a
possible experimental realisation to test the theoretical predictions of our study.}

In this paper, we study the motion of a Brownian particle subjected to Lorentz force with each spatial degree of freedom coupled to a different thermostat. 
We show that the magnetic field gives rise to correlation between different velocity components resulting in spatial cross-correlations. We demonstrate that these correlations persist in the stationary state.
Using a first-principles approach, we calculate the diffusion matrix by integrating the velocity autocorrelation matrix and derive the Fokker-Planck equation for the probability distribution and the corresponding fluxes. \cmnt{The eigenvectors of the diffusion matrix do not align with the temperature axes. As a consequence the Brownian particle performs spatially correlated diffusion.} We show that in contrast to previous studies, even for an isotropic harmonic potential, a nontrivial steady-state density distribution exists, which can be rotated by simply reversing the magnetic field. The steady state, however, is flux-free. By breaking the symmetry in the system using a spatially inhomogeneous magnetic field we show that the Lorentz force induces fluxes in the system. 

The paper is structured as follows. In section~\ref{model}, we provide a description of the model of a diffusion system subjected to Lorentz force with each spatial degree of freedom coupled to a different thermostat. In section~\ref{velocitydistribution}, we derive the conditional probability density of the particle's velocity. We then present the Fokker-Planck equation for the position probability distribution, in section~\ref{diffusionequationsec}. In section~\ref{NESS}, we derive the steady-state solution to the Fokker-Planck equation for the system in an isotropic harmonic potential. Finally, in section~\ref{conclusion}, we present our concluding remarks.

\section{Model}
\label{model}
We consider a single diffusing particle of mass $m$ and charge $q$ subjected to Lorentz force due to an external magnetic field $\B = B \nn$, directed along the unit vector $\nn$. Each spatial degree of freedom of the particle is coupled to a different thermostat at temperature $T_i$ 
where $i=x, y, z$. The stochastic dynamics of the particle are described by the following Langevin equation~\cite{balakrishnan2008elements, vuijk2019anomalous, abdoli2020nondiffusive, abdoli2020stationary}: 
\begin{equation}
m\dot{\vv}(t) = -\gamma \vv + q \vv \times \B + \xxi(t),
\label{langevinequation}
\end{equation}
where $\xxi(t)$ is Gaussian white noise with zero mean and time correlation $\langle\xxi(t)\xxi^{\top}(t')\rangle= 2\gamma k_B\T\delta(t-t')$ where $\gamma$ is the constant friction coefficient and $k_B$ is the Boltzmann constant. Here $\T=\diag(T_x, T_y, T_z)$ is a diagonal  matrix. 

Equation~\eqref{langevinequation} can be rewritten as 
\begin{equation}
	\label{redifinedlangevinequation}
	m\dot{\vv}(t) = -\Gama\cdot\vv(t)+\xxi(t),
\end{equation}
where  $\Gama=\gamma\boldsymbol{1}-qB\M $. 
The elements of the matrix $\M$ are given by $M_{ij}=\epsilon_{ijk}\nn_k$, where $\epsilon_{ijk}$ is the Levi-Civita symbol in three dimensions. 

\cmnt{We also perform Brownian dynamics simulations to validate our analytical predictions. Since the Lorentz force has no effect in the direction of the applied magnetic field, we later restrict our analysis to the motion in the $xy$ plane by applying an external magnetic field in the $z$ direction. Consequently, the simulations are done for a two-dimensional system. In the simulations, the system evolves according to the Langevin equation~\eqref{langevinequation} with a mass $m=0.02$. The chosen mass is sufficiently small for the dynamics to be overdamped. The particle starts its motion from the origin, $(x, y)=(0, 0)$ with the initial velocity $(v_x, v_y)=(1, 1)$. The integration time step is $dt=10^{-4}\tau$. Here, $\tau=\gamma/k_BT_x$, which is the time the particle takes to diffuse over a unit distance.}

\section{Velocity distribution}
\label{velocitydistribution}
Below we outline the procedure to obtain $P(\vv,t|\vv_0)$, which denotes the conditional probability density that a particle with initial velocity $\vv_0$ moves with velocity $\vv$ at time $t$. Equation~\eqref{redifinedlangevinequation} is a linear stochastic differential equation which, using the transformation $\ww(t)=e^{\cmnt{\frac{\Gama}{m}} t}\cdot\vv(t)$, can be written as a Wiener process (see \ref{appendixA} for the details):

\begin{equation}
	\label{wienerequation2}
	m\dot{\ww}(t)=\eeta(t),
\end{equation}
where $\eeta(t)$ is Gaussian white noise with mean $\langle\eeta(t)\rangle=\0$ and time correlation $\langle\eeta(t)\eeta^\top(t')\rangle=2\gamma k_B e^{\cmnt{\frac{\Gama}{m}} t}\cdot\T\cdot e^{\cmnt{\frac{\Gama^\top}{m}} t}\delta(t-t')$, where $\Gama^\top=\gamma\1+qB\M$.

The Fokker-Planck equation corresponding to equation~\eqref{wienerequation2} can be derived using standard methods~\cite{gardiner2009stochastic} and is given as 
\begin{equation}
	\frac{\partial P(\ww, t|\ww_0)}{\partial t}=\nabla_{\ww} \cdot\left[\D_{\ww}(t)\nabla_{\ww}P(\ww, t|\ww_0)\right],
		\label{fokkerplankw}
\end{equation}
where $P(\ww, t|\ww_0)$ is the conditional probability for $\ww$ at time $t$, given that the initial value is $\ww_0$ at time $t=0$. The applied magnetic field is encoded in the matrix $\D_{\ww}(t)$, given by

\begin{equation}
	\label{tensorw}
	\D_{\ww}(t)=\frac{\gamma k_B}{m^2}e^{\cmnt{\frac{\Gama}{m}} t}\cdot\T\cdot e^{\cmnt{\frac{\Gama^\top}{m}} t}.
\end{equation}
\cmnt{Note that the Fokker-Planck equation in \eqref{fokkerplankw} has the same form as that of an inhomogeneous diffusion process. However, the matrix $\boldsymbol{D}_{\boldsymbol{w}}$ is not the matrix for diffusion in position space.}
Also note that for a single-temperature system, $\T = T\1$, the matrix above reduces to a diagonal matrix $\D_{\ww}(t)=e^{\cmnt{\frac{2\gamma}{m}}t}\gamma k_BT/m^2\1$, independent of the magnetic field. 

The fundamental solution to the Fokker-Planck equation~\eqref{fokkerplankw} is the three-dimensional Gaussian distribution in the Cartesian components of $\ww$ which when transformed back to the probability distribution for the velocity (see~\ref{appendixA} for details) reads as
\begin{equation}
\label{solutionv}
\!\!\!\!\!\!\!\!\!\!\!\! P(\vv, t|\vv_0) =
 \frac{\exp\left[-\frac{1}{2}\left(\vv-\uu(t; \vv_0)e^{-\cmnt{\frac{\gamma}{m}}t}\right)^\top\cdot\phii(t)^{-1}\cdot\left(\vv-\uu(t; \vv_0)e^{-\cmnt{\frac{\gamma}{m}}t}\right)\right]}{\sqrt{(2\pi)^3 e^{-\cmnt{\frac{6\gamma}{m}}t}\Det\left(2\int_0^t\D_{\ww}(s)\dif s\right)}},
\end{equation}
where $\uu(t; \vv_0)=e^{\cmnt{\frac{qB\M}{m}}t}\vv_0$ denotes the deterministically evolving initial velocity, modulated by the damping factor $e^{-\cmnt{\frac{\gamma}{m}}t}$. \cmnt{Note that the Lorentz force does not affect the relaxation time of the velocity which remains $m/\gamma$.} The matrix $\phii(t)$ denotes the correlation $\langle (\vv-\uu(t; \vv_0)e^{-\cmnt{\frac{\gamma}{m}}t} )^\top (\vv-\uu(t; \vv_0)e^{-\cmnt{\frac{\gamma}{m}}t}) \rangle$ and is given as 

\begin{equation}
	\label{conditionalvelocityautocorrelation}
	\phii(t)=\frac{2\gamma k_B}{m^2}\int_0^t e^{-\cmnt{\frac{\Gama}{m}} t'}\cdot \T\cdot e^{-\cmnt{\frac{\Gama^\top }{m}}t'}\dif t'.
\end{equation}


We have hitherto considered a multitemperature system subjected to a magnetic field in an arbitrary direction. We now specialize to the case in which the magnetic field points along the $z$ direction with each spatial degree of freedom coupled to a different thermostat. Since the Lorentz force has no effect on the motion along which the magnetic field is pointed (i.e. the $z$ direction), this effectively reduces the problem to a two-dimensional system. As a consequence, we analyze the system in the $xy$ plane. As we show below, this provides an insight into how the Lorentz force affects the steady state properties in a nontrivial fashion. \cmnt{$\phii(t)$ is the velocity autocorrelation function for a specified initial velocity of the Brownian particle. In the long-time limit, the system loses its memory of the initial velocity and attains a steady-state velocity distribution for which the autocorrelation function, denoted by $C(0)$, reads as}



\begin{equation}
\label{conditionalBz}
\CC(0) = \frac{k_B}{2m(1+\kappa^2)} \left( \begin{array}{cccc}
2T_x+\kappa^2(T_x+T_y) & -\kappa(T_x-T_y) \\
-\kappa(T_x-T_y) & 2T_y+\kappa^2(T_x+T_y) \\
\end{array}\right), 
\end{equation}
where the parameter $\kappa=qB/\gamma$ quantifies the strength of the magnetic field relative to frictional force. 

In the equilibrium scenario, i.e., $T_{i} = T$, the velocity autocorrelation reduces to the expected \cmnt{$\delta_{ij}k_BT/m$}, which is independent of the magnetic field. In the general case of different temperatures ($T_x \neq T_y$), there are off-diagonal terms in the matrix which imply cross correlated velocity components in the steady state. Note that these cross-correlations cease to exist in the absence of the magnetic field ($\kappa = 0$).  

 Making a dyadic matrix by multiplying equation~\eqref{redifinedlangevinequation} by $\vv^{\top}(0)$ from the right-hand side, and performing an average over noise and initial velocity (in the steady state), i.e. $\langle\vv(t)\vv^\top(0)\rangle$, we obtain the equation for the time evolution of the velocity autocorrelation $\CC(t)$ as $\dot{\CC}(t) = -\cmnt{\frac{\Gama}{m}} \CC(t)$, which yields
\begin{equation}
	\label{velocityautocorrelation} 
	\CC(t)= e^{-\cmnt{\frac{\Gama}{m}} t} \CC(0),
\end{equation}
where $\CC(0)$ is the initial value of the \cmnt{matrix} in the steady state, which for the special case of magnetic field along the $z$ direction is given in equation~\eqref{conditionalBz}.

\section{Diffusion Equation}\label{diffusionequationsec}

The diffusion equation provides a complete statistical description of the particle's motion in the small-mass limit which corresponds to neglecting inertial effects ($m \rightarrow 0$). This equation is characterized by a diffusion coefficient which in the case of motion under the Lorentz force is a matrix. The matrix encodes the anisotropic nature of diffusion in the presence of a magnetic field. Using the Green-Kubo relation, this matrix can be obtained as an integral of the velocity autocorrelation in equation~\eqref{velocityautocorrelation} as 
\begin{eqnarray}
	\label{kubogreen}
	\D &=& \lim_{t\rightarrow\infty}\int_0^t \CC(t')\dif t',\\
	    &=&\cmnt{m\Gama^{-1}}\CC(0).
\end{eqnarray}

The diffusion equation for the positional probability distribution $P(\rr,t)$ reads as
\begin{equation}
\label{fpe}
\frac{\partial P(\rr,t)}{\partial t} = \nabla \cdot [\D \nabla P(\rr,t)],
\end{equation}
where $\D$ obtained from equations~\eqref{kubogreen} and \eqref{conditionalBz} is given as

\begin{equation}
\D
= \frac{k_B}{\gamma} 
\left( \begin{array}{cccc}
\frac{T_x + \kappa^2 T_y}{\left(1+\kappa^2\right)^2} & -\frac{\kappa(T_x - T_y)}{\left(1+\kappa^2\right)^2}+\frac{\kappa(T_x+T_y)}{2(1+\kappa^2)} \\
-\frac{\kappa(T_x - T_y)}{\left(1+\kappa^2\right)^2}- \frac{\kappa(T_x+T_y)}{2(1+\kappa^2)} & \frac{T_y + \kappa^2 T_x}{\left(1+\kappa^2\right)^2} \\
\end{array}\right).
\label{finaldiffusionmatrix}
\end{equation}

Note that when the magnetic field is absent, equation~\eqref{finaldiffusionmatrix} reduces to the diagonal matrix consistent with the previous studies \cite{dotsenko2013two, mancois2018two}. In the case of the same temperatures along the spatial degrees of freedom and a nonzero magnetic field, the matrix $\D$ reduces to the well-known matrix~\cite{vuijk2019anomalous, abdoli2020nondiffusive, abdoli2020stationary}, which is given as

\begin{equation}
	\label{spatialdiffusionmatrix} 
    \D =\frac{k_BT}{\gamma}\left[\left(\1+\frac{\kappa^2}{1+\kappa^2}\M^2\right)-\frac{\kappa}{1+\kappa^2}\M\right]. 
    \end{equation}

The matrix $\D$ in equation~\eqref{finaldiffusionmatrix} is unusual in the sense that it has antisymmetric part, $\D_a$, and therefore not a typical diffusion matrix. This property of the matrix gives rise to the additional Lorentz fluxes, given as \cmnt{$\J_a = -\D_a\nabla P(\rr,t)$}, which precludes a purely diffusive description with only diffusive fluxes, $\J_s=-\D_s\nabla P(\rr, t)$ even though the underlying dynamics are overdamped~\cite{vuijk2019anomalous, abdoli2020nondiffusive, abdoli2020stationary}. 

The covariance matrix, defined as $\langle \rr(t) \rr^\top(0)\rangle$ is determined only by the symmetric part of $\D$ as $\langle \rr(t) \rr^\top(0)\rangle = 2\D_s t$, where $\D_s = (\D + \D^\top)/2$ is the usual diffusion matrix. 
The variances of the particle's position in the $xy$ plane can be written as

\begin{eqnarray}
\label{variances}
	\label{msdx}
	\langle x^2\rangle &=& \frac{2 k_B(T_x+\kappa^2T_y)}{\gamma(1+\kappa^2)^2}t, \\
	\label{msdy}
	\langle y^2\rangle &=& \frac{2 k_B(T_y+\kappa^2T_x)}{\gamma(1+\kappa^2)^2}t,\\
	\label{mxy}
	 \langle xy\rangle &=& -\frac{2 k_B\kappa(T_x-T_y)}{\gamma(1+\kappa^2)^2}t. 
\end{eqnarray} 

With different temperatures along the $x$ and $y$ axes, it is expected that the motion of the particle is anisotropic. However, since $\langle xy \rangle$ in~\eqref{mxy} is nonzero, the eigenvectors of the diffusion matrix, which are $\nu_1=(\kappa, 1)$ and $\nu_2=(-1/\kappa, 1)$, are not aligned with the temperature axes. The anisotropy in the system can be quantified by the ratio of the eigenvalues, which is simply $T_x/T_y$.  
In the Supplemental Material~\cite{supplements} we show movies of the density distribution and fluxes in the system relaxing towards the equilibrium steady state. The system is initially isotropic by uniformly distributing the particles in a disk. It evolves according to the Fokker-Planck equation in~\eqref{fpe} and becomes anisotropic. The system finally settles into the equilibrium where the density is uniformly distributed.


\begin{figure}
\centering
\resizebox*{0.45\linewidth}{12.5cm}{\includegraphics{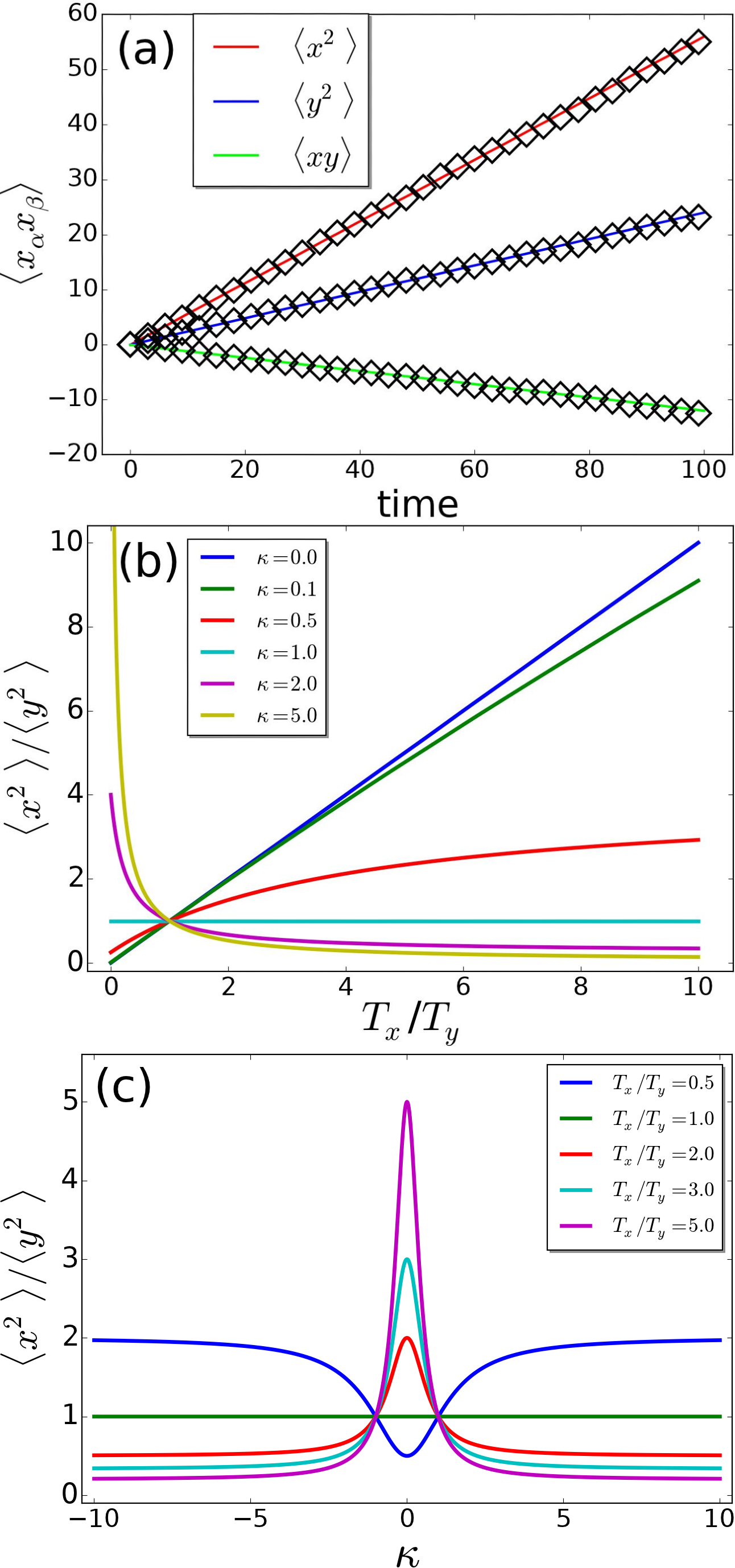}}
\resizebox*{0.45\linewidth}{12.5cm}{\includegraphics{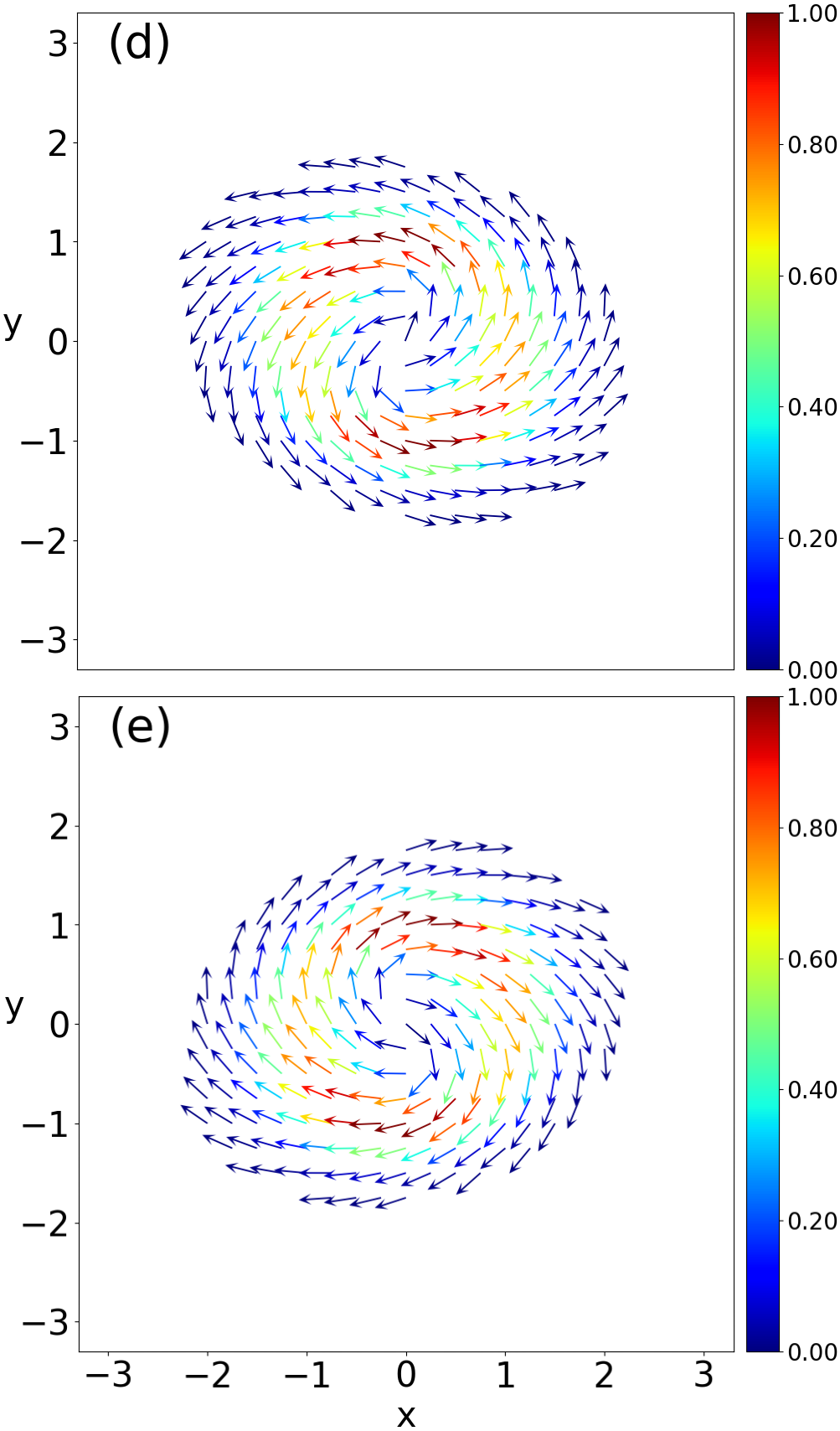}} 
\caption{(a) The variances $\langle x^2\rangle$ and $\langle y^2\rangle$, and the cross correlation $\langle xy\rangle$ are shown by the red, blue, and green lines from the theory~\eqref{msdx}-\eqref{mxy} with $T_y=3T_x=3.0$ and $\kappa=-3.0$. The symbols show the results from the Brownian dynamics simulations. 
The ratio of variances of the displacements of the particle, $\mathcal{R}$, is shown as a function of (b) $T_x/T_y$ for different values of $\kappa$ with $T_y=3T_x=3.0$ and (c) the same as in (b) but as a function of $\kappa$ for different values of $T_x/T_y$. In a system subjected to $\kappa=1.0$ the parameter $\mathcal{R}$ is $1.0$ and independent of $T_x/T_y$. In the case of a single temperature, $T_x/T_y=1.0$ the ratio becomes independent of the applied magnetic field. This parameter reduces to $T_x/T_y$ in the limit that $\kappa\rightarrow 0$, whereas for a large magnetic field , $\kappa\rightarrow\infty$ becomes $T_y/T_x$. \cmnt{The fluxes in system (a) and in a system with an opposite magnetic field are shown in (d) and (e) at time $t = 0.2\tau$, respectively.  The noninteracting particles are initially uniformly distributed in a disk of radius 1 centered at the origin and evolve according to the Fokker-Planck equation~\eqref{fpe}. } The colorbar shows the magnitude of the fluxes which is color-coded. The direction of the fluxes is shown by arrows. } 

\label{varianceratio}
\end{figure}

It is interesting to compare the mean squared displacements along the $x$ and $y$ directions. 
The ratio of the variances, $\mathcal{R}$, is given as 
\begin{equation}
    \label{ratio}
	\mathcal{R}(T_x/T_y; \kappa)\equiv\frac{\langle x^2\rangle}{\langle y^2\rangle} = \frac{\kappa^2+(T_x/T_y)}{1+\kappa^2(T_x/T_y)}.
\end{equation}

In the case of a single temperature corresponding to equilibrium, one obtains $\mathcal{R}=1$ independent of the applied magnetic field. In addition, for a system subjected to $|\kappa|=1.0$, the parameter $\mathcal{R}=1.0$. However, the motion is not isotropic; the anisotropy is encoded in the ratio of the eigenvalues of the diffusion matrix.  In the limit of small magnetic field, $\kappa\rightarrow 0$, the parameter $\mathcal{R}$ reduces to $T_x/T_y$, whereas for a large magnetic field, $\kappa\rightarrow\infty$, approaches $T_y/T_x$.

Figure~\ref{varianceratio} (a) shows the variances of the particle's position as a function of time. The theoretical predictions are in excellent agreement with the Brownian dynamics simulations. 
In figure~\ref{varianceratio}(b) and figure~\ref{varianceratio}(c) we use equation~\eqref{ratio} to plot $\mathcal{R}$ as a function of $T_x/T_y$ for different values of $\kappa$, and $\kappa$ for different values of $T_x/T_y$, respectively. \cmnt{The fluxes in system (a) and in a system with an opposite magnetic field are shown in (d) and (e) at time $t = 0.2\tau$, respectively.  The noninteracting particles are initially uniformly distributed in a disk of radius 1 centered at the origin and evolve according to the Fokker-Planck equation~\eqref{fpe}.}

It is worth considering a somewhat more intuitive approach to the derivation of the diffusion equation. Lets consider a system with temperatures $T_x$ and $T_y$ along the $x$ and $y$ directions, respectively. The motion of \cmnt{the} particle is restricted to the $xy$ plane and the applied magnetic field points in the $z$ direction. Intuitively, one may reason as follows. Consider the flux in $x$ direction. Density gradient along the $x$ direction gives rise to a flux that is proportional to $-T_x\partial_x P$. Due to Lorentz force induced coupling the density gradient along the $y$ direction also contributes to the $x$ flux as $-\kappa T_y\partial_y P$.  With this reasoning, one can now write the components of the flux as $J_x=-\lambda\left(T_x\partial_x+\kappa T_y\partial_y\right)P$ and $J_y=-\lambda\left(T_y\partial_y -\kappa T_x\partial_x \right)P$ where $\lambda^{-1} = \gamma(1 + \kappa^2)k_B^{-1}$ and obtain the diffusion equation as a continuity equation. 
Comparison with equation~\eqref{finaldiffusionmatrix} shows that this intuitive approach is erroneous. In fact, this corresponds to a fictitious system whose diffusion matrix has negative eigenvalues and is therefore unphysical.
The first-principles approach which we present in this work avoids this pitfall.

\section{Nonequilibrium steady state}
\label{NESS}

\begin{figure}
\centering
\resizebox*{0.5\linewidth}{!}{\includegraphics{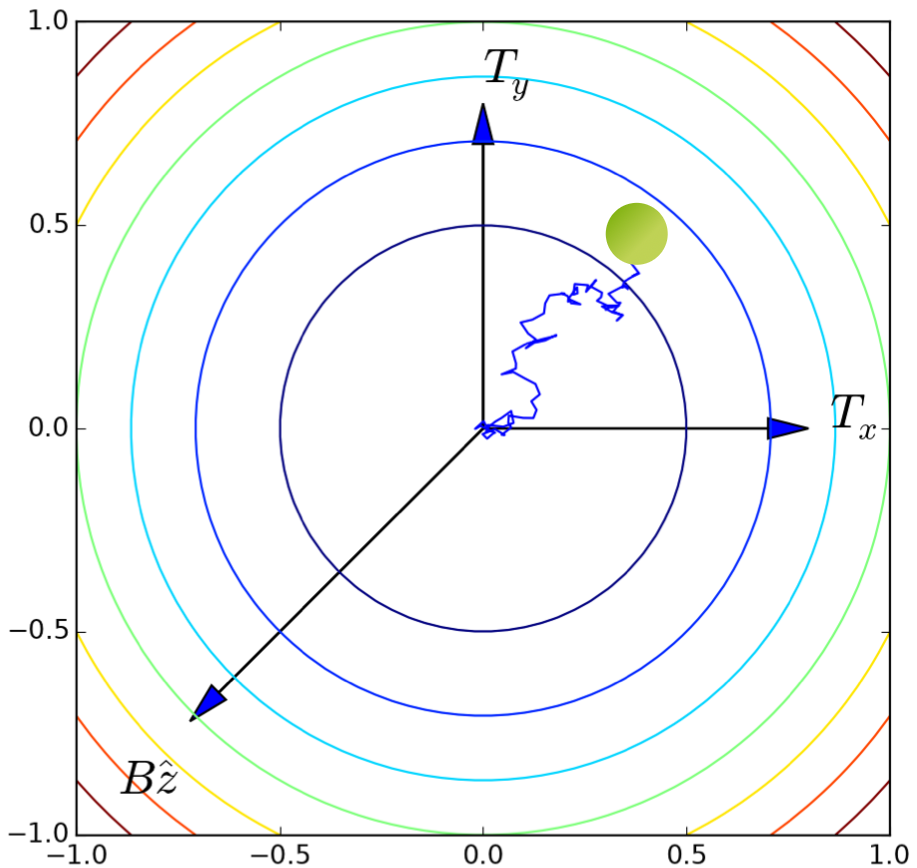}} 
\caption{Two-dimensional Brownian motion subjected to a constant magnetic field along the $z$ direction with different thermostats. The contour plot shows the isotropic harmonic potential $U(x, y)=\varepsilon(x^2+y^2)/2$ with $\varepsilon=2.0$.}

\label{schema}
\end{figure}

In this section, we first use the results derived in section~\ref{diffusionequationsec} to determine the steady state of a particle undergoing Brownian motion in the presence of an \textit{isotropic} harmonic potential $U(x,y)=\varepsilon (x^2 + y^2) /2$ where $\varepsilon$ is the elasticity constant (see figure~\ref{schema}). Then we present our results from simulations for a similar system subjected to a spatially inhomogeneous magnetic field. The Langevin equation governing the dynamics of the particle can be written as:

\begin{equation}
m\dot{\vv}(t) = -\Gama\cdot\vv(t) -\nabla U+ \xxi(t).
\label{langevinequationpotential}
\end{equation}

The Fokker-Planck equation associated with the overdamped equation corresponding to Eq.~\eqref{langevinequationpotential} is given by
\begin{equation}
	\label{diffusionequation}
	\frac{\partial P(x,y, t)}{\partial t} = \nabla\cdot\left[\D\nabla P(x,y, t)+ \Gama^{-1}\nabla U(x,y)P(x,y, t)\right],
\end{equation}
where the matrix $\D$ is given by equation~\eqref{finaldiffusionmatrix}.

\subsection{Uniform magnetic field}
\label{uniformmagneticfield}

For a constant magnetic field, the steady-state solution to equation~\eqref{diffusionequation} is a Gaussian distribution (see~\ref{appendixB} for details), given as

\begin{equation}
\label{probabilitydensity}
P(x, y) = \frac{\varepsilon\sqrt{1+\kappa^2}e^{-(\mu_1x^2+\mu_2y^2+\mu_3xy)}}{\pi\sqrt{\kappa^2(T_x+T_y)^2+4T_xT_y}},
\end{equation}
where

\begin{eqnarray}
     \label{mu1}
	\mu_1 &=& \varepsilon\frac{2T_y + \kappa^2(T_x+T_y)}{\kappa^2(T_x+T_y)^2+4T_xT_y},\\\label{mu2}	
	\mu_2 &=& \varepsilon\frac{2T_x+\kappa^2(T_x+T_y)}{\kappa^2(T_x+T_y)^2+4T_xT_y},\\
	\label{mu3}
	\mu_3 &=& -\varepsilon\frac{2\kappa(T_y-T_x)}{\kappa^2(T_x+T_y)^2+4T_xT_y}.
\end{eqnarray} 

If $\mu_3 \neq  0$, the steady-state probability distribution in equation~\eqref{probabilitydensity} cannot be separated into a product of two independent distributions in $x$ and $y$. Since $\mu_3$ changes sign with $\kappa$, the probability distribution can be rotated by reversing the applied magnetic field. The case $\mu_3 = 0$ corresponds either to (a) an equilibrium system, i.e., $T_x = T_y$, for which, as would be expected, the steady state corresponds to the isotropic equilibrium (Boltzmann) distribution with no dependence on the applied magnetic field or (b) $\kappa = 0$, such that there is no coupling between the spatial degrees of freedom and the distribution is Boltzmann-like. \cmnt{If one changes the handedness of the system from right-handed to left-handed, via the permutations $x \rightarrow y$ and $y \rightarrow x$, it is the same as reversing the applied magnetic field. This is indeed reflected in equations \eqref{mu1} to \eqref{mu3}. Under the permutation, $\mu_1$ and $\mu_2$ remain the same whereas $\mu_3$ changes sign. The same can be obtained in the original right-handed system with reversed magnetic field.}

\begin{figure}
\centering

\resizebox*{0.49\linewidth}{!}{\includegraphics{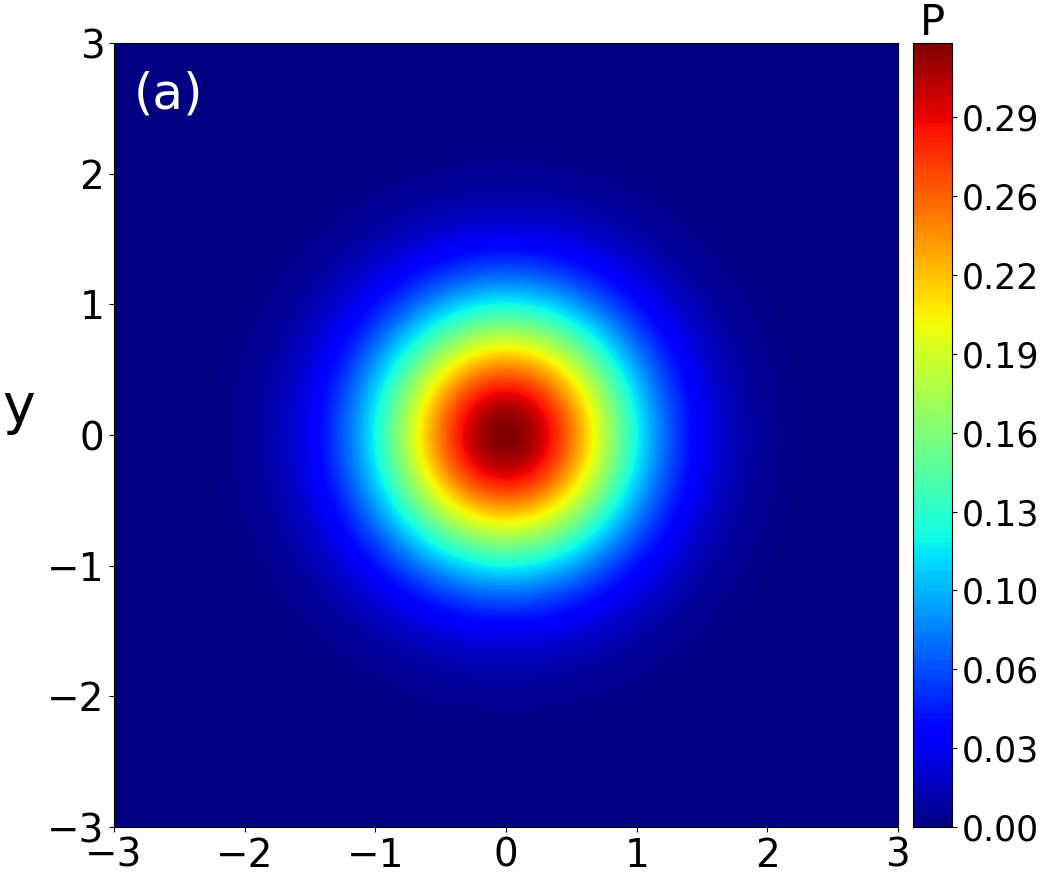}} 
\resizebox*{0.45\linewidth}{!}{\includegraphics{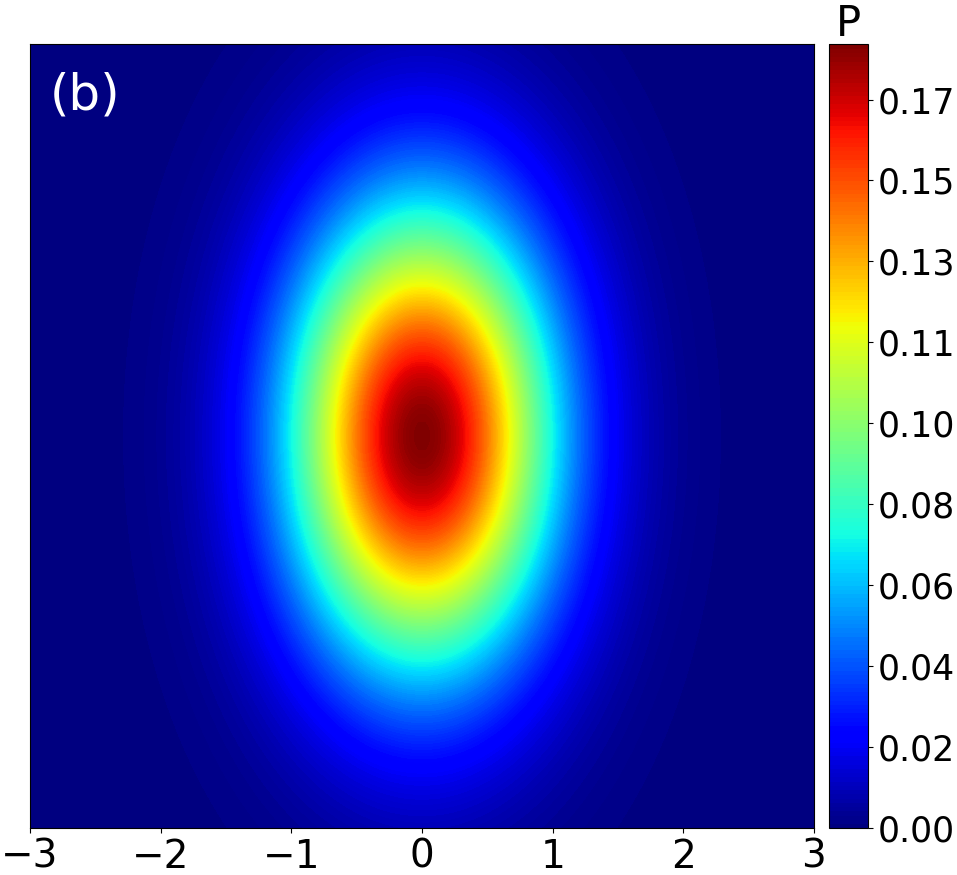}} 
\vspace{0.mm}
\resizebox*{0.49\linewidth}{!}{\includegraphics{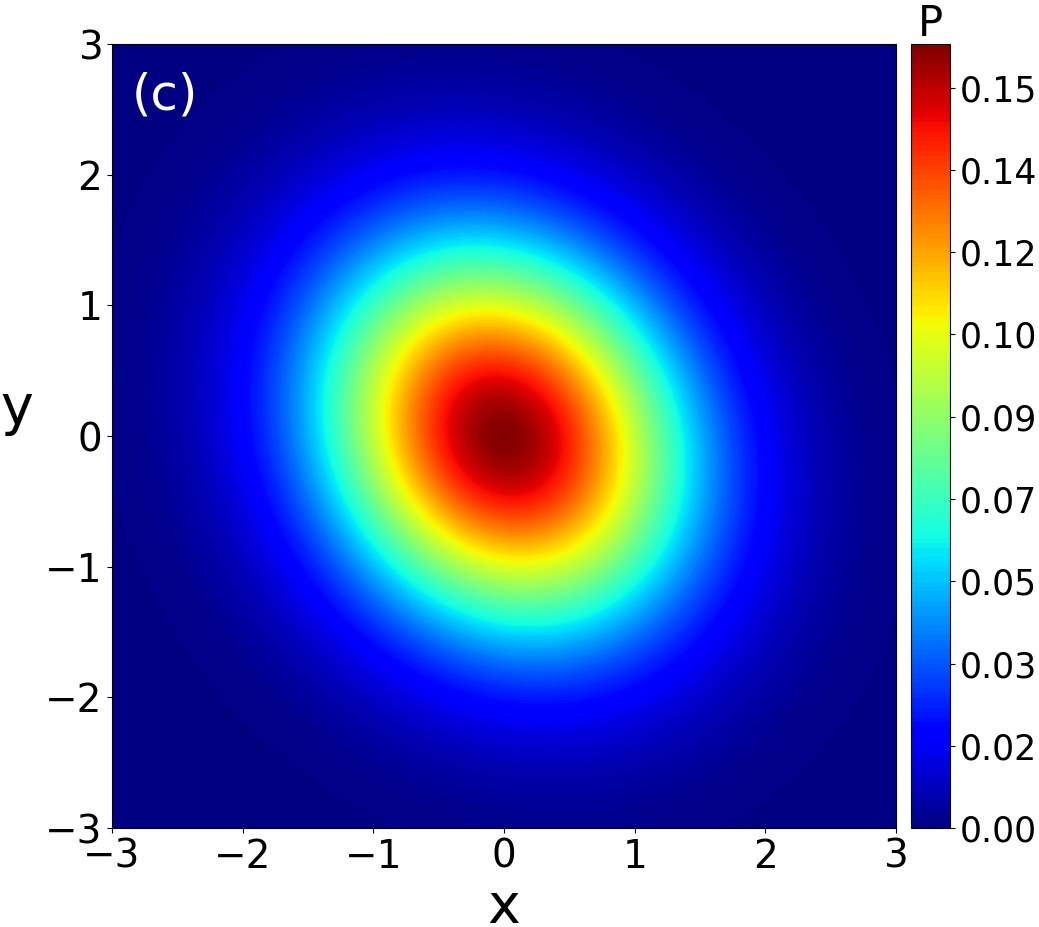}} 
\resizebox*{0.45\linewidth}{!}{\includegraphics{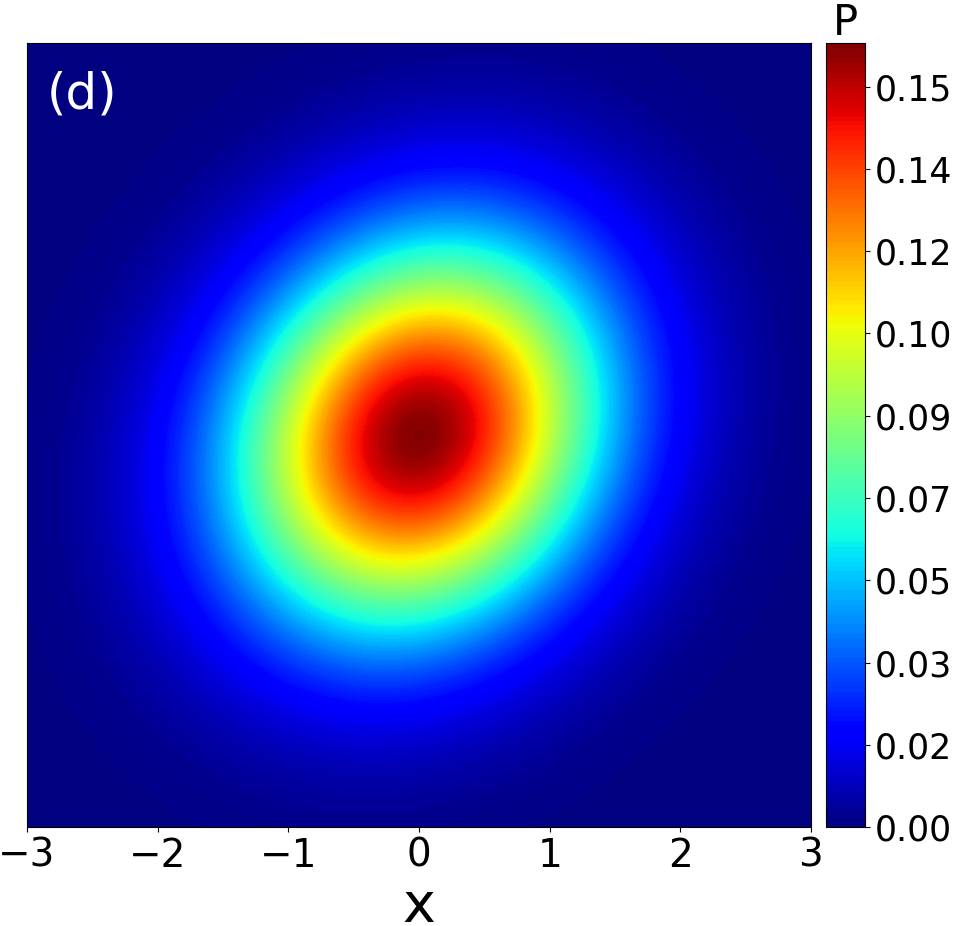}} 
\caption{The stationary probability distribution of the particle's position for the system in an isotropic harmonic potential with $\varepsilon=2.0$. The results are obtained from the analytical predictions in equations~\eqref{probabilitydensity}-\eqref{mu3}. 
(a) corresponds to a single-temperature system $T_x=T_y=1.0$. The applied magnetic field, for the other systems, is constant such that in (b) $\kappa=0.0$, in (c) $\kappa=-3.0$ and in (d) $\kappa=3.0$ with $T_y=3T_x=3.0$. Reversing the direction of the applied magnetic field rotates the density profile by $\pi/2$. \cmnt{A system of two conductors kept at different temperatures and coupled by the electric thermal noise, yields analogous patterns for the joint probability of the voltages where the charges play the role of the positions in the system under study~\cite{ciliberto2013heat}.}}
\label{simulation}
\end{figure}

Figure~\ref{simulation} shows the contour plots of the stationary probability distribution of the position of the particle in (a) a single-temperature system, $T_x=T_y=1.0$, independent of the applied magnetic field  and (b-d) multithermostat systems with $T_y=3 T_x=3.0$ subjected to the constant magnetic fields $\kappa=0.0$, $\kappa=-3.0$, and $\kappa=3.0$, respectively. The elasticity constant is $\varepsilon=2.0$. The probability distribution is rotated by $\pi/2$ on reversing the direction of the applied magnetic field. These results are from the theory in equations~\eqref{probabilitydensity}-\eqref{mu3} and are in full agreement with the simulations (not shown). \cmnt{A system of two conductors kept at different temperatures and coupled by the electric thermal noise, yields analogous patterns for the joint probability of the voltages where the charges play the role of the positions in the system under study~\cite{ciliberto2013heat}.}

Previous studies on two-temperature Brownian systems considered an anisotropic harmonic potential $\varepsilon\left((x^2+y^2)/2 +uxy\right)$~\cite{dotsenko2013two, mancois2018two}, where $u$ is the anisotropy parameter. It was shown that the steady state is nontrivial (not a Boltzmann) only if there exist both anisotropy and temperature difference, captured in the single parameter $u(T_y - T_x)$.  In our system with a (constant) magnetic field, the analogous parameter is $\kappa(T_y - T_x)$ which implies that the $\kappa$ plays a similar role as $u$: it couples $x$ and $y$. 

Despite the magnetic field induced coupling between the spatial degrees of freedom, there are no steady-state fluxes in our system. It is important to note that whereas only the symmetric part of $\D$ enters the calculation for the probability density (see~\ref{appendixB} for details), the flux is calculated using the matrix $\D$ in equation~\eqref{finaldiffusionmatrix}. This flux is zero for the steady-state probability distribution in equation~\eqref{probabilitydensity}. If one took only the symmetric part of $\D$ to calculate the fluxes, one would erroneously conclude that there are steady-state fluxes. We have verified using simulations that the steady state is indeed flux free. \cmnt{The absence of steady-state fluxes in our system is in contrast to previous studies of multithermostat systems~\cite{dotsenko2013two, mancois2018two} in which fluxes existed in the stationary state.}   

\subsection{Spatially inhomogeneous magnetic field}
\label{inhomogeneousmagneticfield}
\begin{figure}
\centering
\resizebox*{0.5\linewidth}{!}{\includegraphics{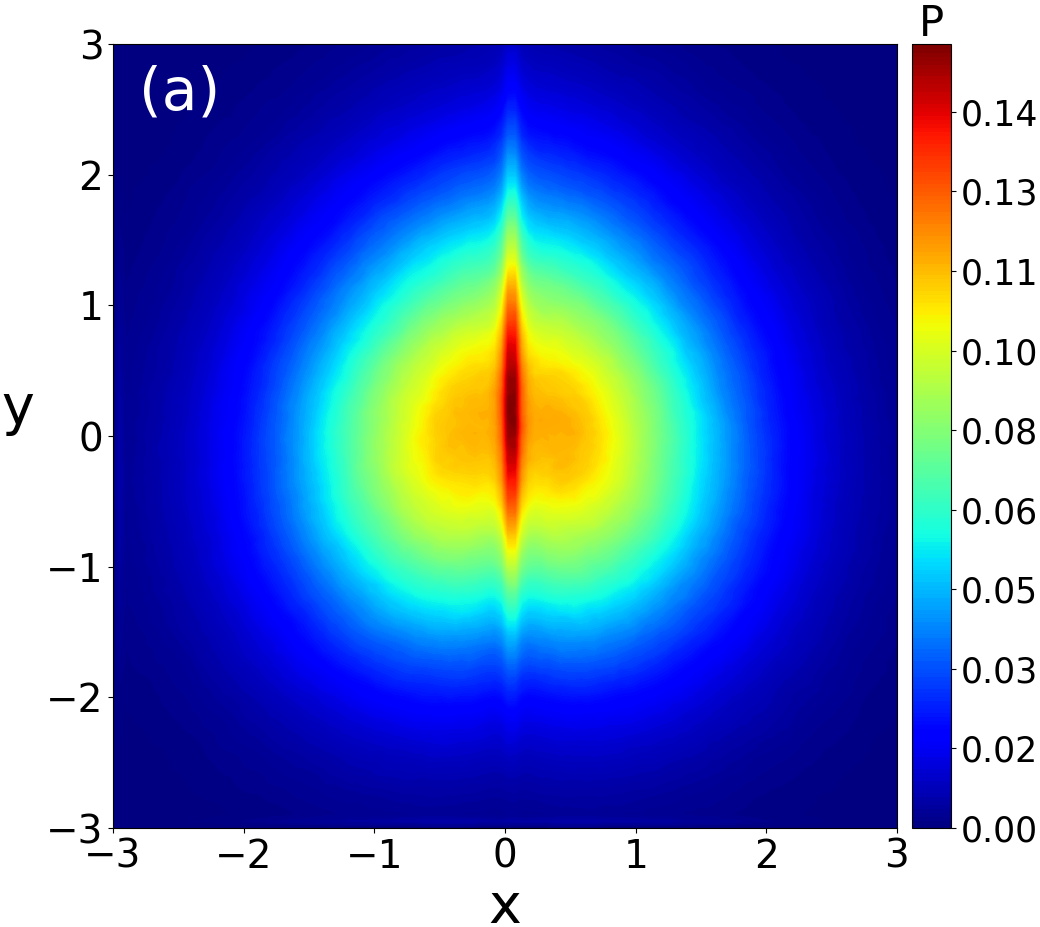}} 
\resizebox*{0.46\linewidth}{!}{\includegraphics{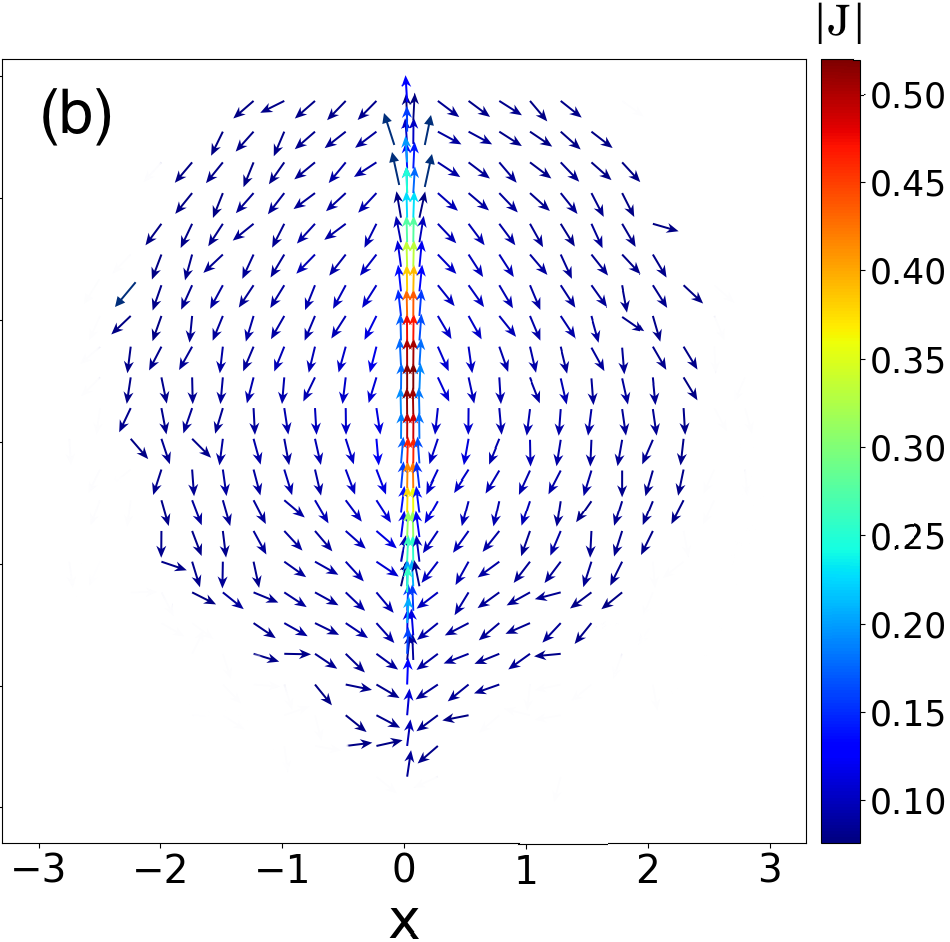}} 

\caption{(a) The contour plot of the stationary probability distribution of the particle's position and (b) the fluxes for the system in an isotropic harmonic potential with $\varepsilon=2.0$, and $T_y= 3T_x=3.0$. The system is divided into two half-planes by the line $x=0$. Each half plane is subjected to a constant magnetic field such that $\kappa=-3.0$ if $x>0$ and $\kappa=3.0$ otherwise. The distribution is highly stretched along the line $x=0$. The direction of the fluxes is shown by the arrows; the magnitude is color coded.}

\label{inhomogeneousBfig}
\end{figure}

\cmnt{In the previous studies, it was shown that the emergence of the fluxes is due to two broken symmetries: a different temperature in each spatial degree of freedom and a mismatch between the principal axes of temperature and those of the potential. Here we show that even in a system with an isotropic potential, fluxes may be induced by breaking symmetry in the system via an inhomogeneous magnetic field.} In addition to the broken symmetry of two different temperatures along the spatial axes, we break another symmetry in the system by using a spatially inhomogeneous magnetic field. We divide the system into two half-planes at the line $x=0$. Each half-plane is subjected to a constant magnetic field with the same magnitude, but opposite direction such that $\kappa=-3.0$ for $x>0$ and $\kappa=3.0$ otherwise. Symmetry requires that (a) the steady-state probability distribution is even in $x$ and (b) zero flux in the $x$ direction at $x = 0$. Due to the applied magnetic field, the probability distribution is not symmetric in $y$ implying that $\langle y \rangle \neq 0$. Figure~\ref{inhomogeneousBfig} (a) shows a contour plot of the stationary probability distribution obtained from Brownian dynamics simulations of equation~\eqref{langevinequationpotential}. As expected, the average position of the particle is displaced from the origin along the $y$ axis.

In contrast to the constant magnetic field case, there are fluxes in the nonequilibrium steady state as shown In figure~\ref{inhomogeneousBfig} (b). As can be seen, the $x$ component of the fluxes is zero at $x=0$. 
\section{Concluding Remarks}
\label{conclusion}

Since the Lorentz force due to a magnetic field performs no work, it does not affect equilibrium properties of a system. It does, however, give rise to dynamics which are fundamentally different from a purely diffusive system~\cite{chun2018emergence,abdoli2020nondiffusive}. It can generate unusual nonequilibrium steady states which are quite distinct from those generated by other non-conservative driving forces (e.g. shear), which input energy to the system. The stationary state is generally characterized by a non-Boltzmann density distribution and fluxes~\cite{vuijk2019lorenz,abdoli2020stationary}.

In the presence of the Lorentz force, correlations appear in the velocity due to the mixing of different velocity components. Whereas the correlations are transient in a single-temperature (equilibrium) system, in a multithermostat system different velocity components remain correlated in the stationary state. As a consequence, the spatial degrees of freedom  remain correlated in the stationary state giving rise to an anisotropic diffusion matrix with its eigenvectors misaligned with those of the temperature. The Lorentz force induced coupling is quite distinct from previous studies of multithermostat systems (without Lorentz force) in which the spatial degrees of freedom were coupled via an anisotropic potential~\cite{dotsenko2013two, mancois2018two}. Spatial correlations exist only when the principal axes of the potential do not match with the temperature axes.

In this paper, we showed that on confining the particle via an isotropic harmonic potential, an interesting stationary state emerges: it has a nontrivial density distribution that depends on the applied magnetic field but is otherwise flux-free. However, by breaking the symmetry in the system using a spatially inhomogeneous magnetic field, the Lorentz force induces fluxes in the system. \cmnt{For $T_x = T_y$ the magnetic field contribution drops out entirely from the equilibrium distribution. This is in fact expected from the Bohr-van Leeuwen theorem \cite{vanleeuwen:jpa-00204299, roth1967plasma} which states that the thermal average of magnetization is always zero in an equilibrium system.}

In future work, we will extend the idea of the current study to include interacting particles~\cite{hosaka2017thermally,sou2019nonequilibrium}. It would also be interesting to study the escape problem for a multithermostat system~\cite{hanggi1990reaction,sharma2017escape, chupeau2020optimizing,scacchi2018mean} . Moreover one could study whether some of the phenomenology in the multithermostat system can be reproduced by stochastically resetting the particle to the axes with different rates. 

\appendix 
\section{Derivation of velocity autocorrelation} \label{appendixA}
Here, we first rewrite equation~\eqref{langevinequation} as a Wiener process and derive the corresponding Fokker-Planck equation. Then, we solve the resulting diffusion equation and perform an inverse transformation to obtain the solution in the velocity space. 
We start with equation~\eqref{redifinedlangevinequation} and multiply both sides by the integrating factor $e^{\cmnt{\frac{\Gama}{m}} t}$, which yields
\begin{equation}
	\label{langevinapp}
	\frac{\dif}{\dif t}\left[e^{\cmnt{\frac{\Gama}{m}} t}\cdot\vv(t)\right]=\frac{1}{m}e^{\cmnt{\frac{\Gama}{m}} t}\cdot\xxi(t).
\end{equation}

The variable transformation $\ww(t)=e^{\cmnt{\frac{\Gama}{m}} t}\cdot\vv(t)$ turns this equation in a Wiener process:
\begin{equation}
	\label{wienerapp}
	\dot{\ww}(t)=\frac{1}{m}\eeta(t),
\end{equation}  
where $\eeta(t)$ is a new stochastic noise with 
\begin{eqnarray}
	\langle\eeta(t)\rangle & = & \0 \label{meanapp},\\
	\langle\eeta(t)\eeta^\top(t')\rangle & = & 2\gamma k_B e^{\cmnt{\frac{\Gama}{m}} t}\cdot\T\cdot e^{\cmnt{\frac{\Gama^\top}{m}} t}\delta(t-t'). \label{correlationapp} 
\end{eqnarray}
The Fokker-Planck equation corresponding to equation~\eqref{wienerapp} can be obtained using standard methods~\cite{gardiner2009stochastic} which reads as
\begin{equation}
	\label{fpeapp}
	\frac{\partial P(\ww, t|\ww_0)}{\partial t}=-\nabla_{\ww}\cdot\jj(\ww, t|\ww_0),
\end{equation}
where $\jj(\ww, t|\ww_0)$ is the flux in the $\ww$ space, given as
\begin{equation}
	\label{fluxapp}
	\jj_w(\ww, t|\ww_0)= -\D_{\ww}\nabla_{\ww}P(\ww, t|\ww_0), 
\end{equation}
where $\D_{\ww}(t)=\langle\eeta(t)\eeta^\top(t)\rangle/2m^2$. The substitution of equation~\eqref{fluxapp} into equation~\eqref{fpeapp} results in equation~\eqref{fokkerplankw} in the main text. Using the Fourier transformation equation~\eqref{fokkerplankw} can be solved. The transformed equation can be written as

\begin{equation}
	\label{transformedfpeapp}
	\frac{\partial\tilde{P}(\kk, t)}{\partial t}=-\left[\kk^\top\cdot\D_{\ww}(t)\cdot\kk\right]\tilde{P}(\kk, t).
\end{equation} 
where the tilde indicates a Fourier transformation from the variable $\ww$ into $\kk$. The solution to equation~\eqref{transformedfpeapp} reads


\begin{equation}
\label{transformedsolutionapp}
\tilde{P}(\kk, t) =
 \exp\left[-\frac{1}{2}\kk^\top\cdot\left(2\int_0^t\D_{\ww}(s)\dif s\right)\cdot\kk\right].
\end{equation} 
The inverse Fourier transform of equation~\eqref{transformedsolutionapp} is

\begin{equation}
\label{solutionw}
P(\ww, t|\ww_0) = \frac{
 \exp\left[-\frac{1}{2}(\ww-\ww_0)^\top\cdot\left(2\int_0^t\D_{\ww}(s)\dif s\right)^{-1}\cdot(\ww-\ww_0)\right]}{\sqrt{(2\pi)^3\Det\left(2\int_0^t\D_w(s)\dif s\right)}},
\end{equation}

In order to obtain the probability distribution of the velocity of the particle we use the following transformation
\begin{equation}
	\label{jacobiantransformationapp}
	P(\vv, t|\vv_0)= \mathrm{J}_{\ww}(\vv, t)P(\ww,t|\ww_0),
\end{equation}
where 
\begin{eqnarray}
\label{jacobianapp1}
	\mathrm{J}_{\ww}(\vv,t)& = & \Det\left(\frac{\partial(w_1, w_2, w_3)}{\partial(v_1,v_2,v_3)}\right), \\
	                 & = & e^{\tr\left[\cmnt{\frac{1}{m}}(\gamma\1-qB\M)t\right] },
\end{eqnarray}
is the Jacobian reflecting the variable transform from $\ww$ to $\vv$. The trace of the matrix $\M$ is zero which results in $\mathrm{J}_{\ww}(\vv,t)=e^{\cmnt{\frac{3\gamma}{m}} t}$
when plugged into equation~\eqref{jacobiantransformationapp} gives
\begin{equation}
\label{velocitysolutionapp}
\!\!\!\!\!\!\!\!\!\!\!\!
\tilde{P}(\vv, t|\vv_0) = \frac{
 \exp\left[-\frac{1}{2}(e^{\cmnt{\frac{\Gama}{m}} t}\vv-\vv_0)^T\cdot\left(2\int_0^t\D_{\ww}(s)\dif s\right)^{-1}\cdot(e^{\cmnt{\frac{\Gama}{m}} t}\vv-\vv_0)\right]}{{\sqrt{(2\pi)^3e^{-\cmnt{\frac{6\gamma}{m}} t}\Det\left(2\int_0^t\D_w(s)\dif s\right)}}},
\end{equation}
where $\vv_0=\ww_0$. Considering $e^{\cmnt{\frac{\gamma\1}{m}} t}=e^{\cmnt{\frac{\gamma}{m}} t}\1$ and $\left(e^{-\cmnt{\frac{qB\M}{m}} t}\right)^\top=e^{\cmnt{\frac{qB\M}{m}} t}$, equation~\eqref{velocitysolutionapp} results in equation~\eqref{solutionv} in the main text, where the conditional velocity autocorrelation is defined as
\begin{equation}
	\label{velocityautoapp1}
	\phii(t)=\left[\left(e^{\cmnt{\frac{\Gama}{m}} t}\right)^\top\cdot\left(2\int_0^t\D_{\ww}(s)\dif s\right)^{-1}\cdot e^{\cmnt{\frac{\Gama}{m}} t}\right]^{-1}.
\end{equation}
By substituting $\D_w$, defined after equation~\eqref{fluxapp}, into equation~\eqref{velocityautoapp1} one gets
\begin{equation}
	\label{velocityautoapp2}
	\phii(t)=\frac{2\gamma k_B}{m^2}\int_0^t e^{-\cmnt{\frac{\Gama}{m}}(t-s)}\cdot\T\cdot e^{-\cmnt{\frac{\Gama^\top}{m}}(t-s)}\dif s,
\end{equation}
Making the change in variables $t'=t-s$ 
, equation~\eqref{velocityautoapp2} can be written as
\begin{equation}
	\label{velocityautoapp3}
	\phii(t)=\frac{2\gamma k_B}{m^2}\int_0^t e^{-\cmnt{\frac{\Gama}{m}} t'}\cdot \T\cdot e^{-\cmnt{\frac{\Gama^\top}{m}} t'}\dif t'.
\end{equation}
The derivation of equation~\eqref{conditionalvelocityautocorrelation} is complete. It can be alternatively represented in terms of the the eigenvalues and eigenvectors of the matrices $\Gama$ and $\Gama^\top$, given as
\begin{eqnarray}
	\label{eigenapp1a}
	\Gama\LL &=&\cmnt{\LL\Lambdaa,} \\
	\label{eigenapp1b}
	\Gama^\top\R &=&\cmnt{\R\Lambdaa,}
\end{eqnarray}
where $\Lambdaa$ is a diagonal matrix of the eigenvalues of the matrix $\Gama$ (and $\Gama^\top$), and $\LL$ (and $\R$) is the corresponding matrix of eigenvectors. Using equation~\eqref{eigenapp1a} and equation~\eqref{eigenapp1b} one can write
\begin{eqnarray}
	\label{eigenapp2a}
	e^{-\cmnt{\frac{\Gama}{m}} t'} &=&\LL e^{-\cmnt{\frac{\Lambdaa}{m}} t'}\LL^{-1}, \\
	\label{eigenapp2b}
	e^{-\cmnt{\frac{\Gama^\top}{m}} t'} &=&\R e^{-\cmnt{\frac{\Lambdaa}{m}} t'}\R^{-1}.
\end{eqnarray}

Plugging equation~\eqref{eigenapp2a} and equation~\eqref{eigenapp2b} into equation~\eqref{velocityautoapp3}, one gets

\begin{equation}
	\label{velocityautoapp4}
	\phii(t) = \frac{2\gamma k_B}{m^2}\LL\left(\int_0^t e^{-\cmnt{\frac{\Lambdaa}{m}} t'}\hat{\T}e^{-\cmnt{\frac{\Lambdaa}{m}} t'}\dif t'\right)\R^{-1},
\end{equation}
where $\hat{\T}=\LL^{-1}\T\R$ can be interpreted as the matrix of rotated and mixed temperatures due to the magnetic field.

\section{Derivation of the steady-state solution}
\label{appendixB}
In this section, we present the method which is used to obtain the solution to equation~\eqref{diffusionequation} which, for convenience, we recall it:

\begin{equation}
	\label{diffusionequationappendix}
	\frac{\partial P(x,y, t)}{\partial t} = \nabla\cdot\left[\D\nabla P(x,y, t)+\Gama^{-1}\nabla U(x,y)P(x,y, t)\right],
\end{equation}

This equation is a linear multivariate Fokker-Planck equation with
a Gaussian solution, given as~\cite{mancois2018two}
\begin{equation}
	\label{gaussiandistribution}
	P(x, y, t)=\cmnt{\frac{1}{2\pi\sqrt{\Det(\boldsymbol{X})}}}\exp\left[-\frac{1}{2}(x, y)^\top\cdot\boldsymbol{X}^{-1}\cdot(x, y)\right],
\end{equation}
where $\boldsymbol{X}$ is the covariance matrix which satisfies the following Lyapunov equation:
\begin{equation}
	\label{liapunovequation}
	\frac{\dif\boldsymbol{X}}{\dif t}=\A\boldsymbol{X}+\boldsymbol{X}\A^\top+\B,
\end{equation}
where the matrix $\A\equiv-\varepsilon\Gama^{-1}$ is given by

\begin{equation}
\A 
= -\frac{\varepsilon}{\gamma(1+\kappa^2)} 
\left( \begin{array}{cccc}
1 & \kappa  \\
-\kappa & 1  \\
\end{array}\right),
\label{matrixA} 
\end{equation} 
and the matrix $\B\equiv 2\D_s$ is given as

\begin{equation}
\B
= \frac{2 k_B}{\gamma(1+\kappa^2)^2} 
\left( \begin{array}{cccc}
T_x + \kappa^2 T_y & -\kappa(T_x-T_y)  \\
-\kappa(T_x-T_y) & T_y + \kappa^2 T_x  \\
\end{array}\right),
\label{matrixB}
\end{equation} 
For the stationary solution the stationary covariance matrix, $\boldsymbol{X}_{SS}$ obeys the corresponding Lyapunov equation by setting $\dif\boldsymbol{X}/\dif t=\0$. This implies that the solution to the stationary Lyapunov equation reads
\begin{equation}
	\label{genericsoltion}
	\boldsymbol{X}_{SS}=\int_0^\infty e^{\A t}\cdot\B\cdot e^{\A^\top t}\dif t.
\end{equation}

By plugging equation~\eqref{matrixA} and equation~\eqref{matrixB} into equation~\eqref{genericsoltion}, one gets the covariance matrix as 

\begin{equation} 
\boldsymbol{X}_{SS} 
= \frac{1}{2\varepsilon k_B(1+\kappa^2)} 
\left( \begin{array}{cccc}
2T_x+\kappa^2(T_x+T_y) & \kappa(T_x-T_y)  \\
\kappa(T_x-T_y) & 2T_y + \kappa^2(T_x+T_y)  \\
\end{array}\right).
\label{covariancematrix}
\end{equation}

The substitution of equation~\eqref{covariancematrix} into equation~\eqref{gaussiandistribution} 
results in the solution to the corresponding Fokker-Planck equation~\eqref{diffusionequationappendix} which is given by equation~\eqref{probabilitydensity} in the main text. 

\section*{References}
\providecommand{\noopsort}[1]{}\providecommand{\singleletter}[1]{#1}%
\providecommand{\newblock}{}

\end{document}